\documentclass[preprint2]{proto}
\usepackage{times}

\newcommand{\refs}{\par\noindent\hangindent=1pc\hangafter=1}
\begin{document}

\title{\textbf{\LARGE New Observational Frontiers in the Multiplicity
    of Young Stars}}

\author {\textbf{\large Gaspard Duch\^ene}}
\affil{Laboratoire d'Astrophysique de Grenoble}
\author {\textbf{\large Eduardo Delgado-Donate}}
\affil{Stockholm Observatory}
\author {\textbf{\large Karl E. Haisch Jr.}}
\affil{Utah Valley State College}
\author {\textbf{\large Laurent Loinard and Luis F. Rodr\'{\i}guez}}
\affil{Universidad Nacional Aut\'onoma de M\'exico}

\begin{abstract}
\baselineskip = 11pt
\leftskip = 0.65in
\rightskip = 0.65in
\parindent=1pc {\small It has now been known for over a decade that
  low-mass stars located in star-forming regions are very frequently
  members of binary and multiple systems, even more so than main
  sequence stars in the solar neighborhood. This high multiplicity
  rate has been interpreted as the consequence of the fragmentation of
  small molecular cores into a few seed objects that accrete to their
  final mass from the remaining material and dynamically evolve into
  stable multiple systems, possibly producing a few ejecta in the
  process. Analyzing the statistical properties of young multiple
  systems in a variety of environments therefore represents a powerful
  approach to place stringent constraints on star formation theories.
  In this contribution, we first review a number of recent results
  related to the multiplicity of T~Tauri stars. We then present a
  series of studies focusing on the multiplicity and properties of
  optically-undetected, heavily embedded protostars. These objects are
  much younger than the previously studied pre-main sequence stars,
  and they therefore offer a closer look at the primordial population
  of multiple systems. In addition to these observational avenues, we
  present new results of a series of numerical simulations that
  attempt to reproduce the fragmentation of small molecular cores into
  multiple systems, and compare these results to the observations.
  \\~\\~\\~}

\end{abstract}

\section{\textbf{INTRODUCTION}}

The prevalence of binary and higher-order multiple systems is a
long-established observational fact for field low-mass stars ({\it
Duquennoy and Mayor}, 1991; {\it Fischer and Marcy}, 1992). For over a
decade, it has been known that young pre-main sequence stars are also
often found in multiple systems. The chapter by {\it Mathieu et al.}
(2000) in the previous volume in this series has summarized various
statistical surveys for visual multiple systems among T~Tauri stars in
star-forming regions as well as of zero-age main sequence stars in
open clusters. These early multiplicity surveys have shown that
multiple systems are ubiquitous among young stellar objects (YSOs) and
further revealed an environment-dependent trend. The multiplicity rate
in all stellar clusters, even those with the youngest ages such as the
Orion Trapezium cluster, is in excellent agreement with that observed
in main sequence field stars. On the other hand, the least dense
T~Tauri populations, like the Taurus-Auriga and Ophiuchus clouds, show
a factor of $\sim$2 multiplicity excess, relative to field low-mass
stars. However, it remained impossible to decide whether this behavior
was the consequence of an intrinsic difference in the fragmentation
process, or the result of dynamical disruptive interactions acting on
timescales shorter than 1~Myr in stellar clusters (see {\it Patience
and Duch\^ene}, 2001 for a review).

The purpose of this chapter is to review a variety of
observational results concerning the multiplicity of young low-mass
stars in order to update the view presented by {\it Mathieu et al.}
(2000). In addition, we present some numerical results related to the
fragmentation and subsequent evolution of low-mass prestellar
cores. These models make predictions that can be readily tested with
the observational results discussed here. Throughout this chapter, we
focus on low-mass stellar objects with masses roughly ranging from
0.1 to 2$M_\odot$. The multiplicity of young substellar objects is
discussed in detail in the chapters by {\it Burgasser et al.} and
{\it Luhman et al}, whereas the multiplicity of higher-mass objects is
addressed in the chapter by {\it Beuther et al.} Other chapters in
this volume present complementary insights on the subject: {\it
Goodwin et al.} present more numerical results on the collapse and
fragmentation of molecular cores, as well as on the dynamical evolution
of small stellar aggregates; {\it Whitworth et al.}, {\it
Ballesteros-Paredes et al.} and {\it Klein et al.}) discuss the
collapse of larger-scale molecular cores; {\it Mathieu et al.} and
{\it Monin et al.}  focus on various properties (dynamical masses and
disks properties, respectively) of known T~Tauri binary stars.
\bigskip

\noindent{\textbf{ 2. AN UPDATE ON THE MULTIPLICITY OF YOUNG LOW-MASS
    STARS}} 
\bigskip

As mentioned above, the first efforts to study young multiple systems
were focused on determining the average number of wide companions per
object in well-known pre-main sequence populations. Attempting to
account for these multiplicity rates has led to various theories that
involve the fragmentation of molecular cores and the subsequent
dynamical evolution of aggregates of stars embedded in gaseous
clouds. So far, the observed multiplicity rate of T~Tauri populations
alone has not proved entirely conclusive, and since the review by {\it
Mathieu et al.} (2000), the focus of statistical studies of young
multiple systems has shifted to other areas. Before probing much
younger, still embedded multiple systems (Section~3), we review here a
number of studies on T~Tauri multiple systems that go beyond the
surveys that were conducted in the 1990s.

\bigskip
\noindent
\textbf{ 2.1 Multiplicity In Young Nearby Associations}
\bigskip

The clear dichotomy between high- and low-multiplicity star-forming
regions has usually been considered evidence of an
environment-dependent star formation scenario. However, since most
stars form in stellar clusters, one could also consider that the rare
molecular clouds that host too many companions are exceptions for yet
undetermined reasons. Over the last decade, several groups of a few
tens of stars with ages typically between $\sim$10 and 50~Myr have
been identified in the Sun's vicinity based on their common
three-dimensional motion and youth indicators ({\it Zuckerman and
  Song}, 2004). Most members of these associations are low-mass
pre-main sequence stars. Therefore, these co-moving groups represent
additional, nearby populations of young stars whose multiplicity could
be expected to resemble that of the Taurus-Auriga population given
their low stellar densities.

Soon after their discovery, systematic searches for visual companions
were conducted in some of these groups in order to complement the
previous surveys. For instance, {\it Chauvin et al.} (2002, 2003) and
{\it Brandeker et al.} (2003) conducted surveys for visual companions
in the TW~Hya, Tucana-Horologium and MBM~12 groups; we include MBM~12
in this discussion despite the continuing debate regarding its
distance ($\sim$65~pc according to {\it Hearty et al.}, 2000, revised
upwards to $\sim275$~pc by {\it Luhman}, 2001). In their review of all
known nearby associations, {\it Zuckerman and Song} (2004) marked
those systems that were discovered to be multiple in these surveys or
during pointed observations of individual objects. The average
multiplicity rate in these associations range from 20\% to over 60\%
but the small sample sizes preclude clear conclusions on any
individual association. Averaging all associations listed in {\it
Zuckerman and Song}'s review, and adding the MBM~12 surveys from {\it
Chauvin et al.}  (2002) and {\it Brandeker et al.}  (2003), the
average number of visual companions per member is
38.2$\pm$3.6\%. Because of the range of distances to the stars
involved in this survey, it is difficult to compare this to previous
surveys of T~Tauri stars, which were surveyed over a more homogeneous
separation range. Nonetheless, the observed multiplicity rate in
nearby young associations appears to be high, possibly as high as in
Taurus-like populations. Future dedicated studies sampling a uniform
separation range will help reinforce this conclusion.

\bigskip
\noindent
\textbf{ 2.2 Multiplicity Of The Lowest Mass T~Tauri Stars}
\bigskip

Most of the T~Tauri multiplicity surveys summarized in {\it Mathieu et
al} (2000) were focusing on objects with masses in the range
0.5--2$M_\odot$, essentially because of the limited sensitivity of
high-angular resolution devices at that time. Multiplicity surveys
conducted in recent years have, therefore, focused primarily on the
multiplicity rate of the lowest mass T~Tauri stars in known
star-forming regions in order to determine the mass-dependency of the
properties of multiple systems.

We first focus on systematic surveys for multiplicity among low-mass
(0.1--0.5$M_\odot$) T~Tauri stars. {\it White et al.} (in prep.) have
obtained high angular resolution datasets on $\sim50$ such objects in
Taurus-Auriga that represent a nice comparison sample to the early
surveys of {\it Ghez et al} (1993) and {\it Leinart et al.} (1993),
for instance. They find that even low-mass T~Tauri stars have a high
multiplicity rate, although with a decreasing trend towards the lowest
stellar masses. In addition, they find that multiple systems in which
the primary has a mass $M_A\lesssim0.4M_\odot$ are confined to mass
ratios higher than $M_B/M_A\sim0.6$, and very rarely have projected
separations larger than $\sim200$ AU despite sensitive searches. Apart
from the overall multiplicity excess among T~Tauri stars, these trends
are in line with the results of multiplicity surveys among lower mass
main sequence field stars ({\it Marchal et al.}, 2003; {\it Halbwachs
  et al.}, 2003). These various mass-dependencies must be explained by
models of fragmentation and early evolution of multiple systems.

Extending this approach to and beyond the substellar limit, {\it Bouy
et al.} (2003) in the Pleiades and {\it Kraus et al.} (2005) in Upper
Scorpius found that the trends for lower metalicity, tighter, and
preferentially equal-mass systems are amplified in the brown dwarf
regime, although the exact multiplicity frequency of young brown
dwarfs is still under debate. Pre-main sequence brown dwarfs are
discussed in more detail elsewhere in this volume (see chapters
by {\it Luhman et al.} and {\it Burgasser et al.}).

\bigskip
\noindent
\textbf{ 2.3 Dynamical Masses Of Binary T~Tauri Systems}
\bigskip

The masses of T~Tauri stars are usually determined through their
location in the HR diagram in comparison to pre-main sequence
evolutionary models. However, there is a long-standing debate on the
validity domain of these models, which all include some, but not all,
of the key physical ingredients (e.g., {\it Baraffe et al.},
2002). Empirical mass determinations for T~Tauri stars have been
attempted for many years, largely through the study of binary and
multiple systems. A thorough analysis of the confrontation of current
evolutionary models with empirical mass determinations of T~Tauri
stars has recently been presented by {\it Hillenbrand and White}
(2004).

There are only a handful of known pre-main sequence eclipsing binaries
({\it Mathieu et al.}, 2000; {\it Covino et al.}, 2000, 2004; {\it
Stassun et al.}, 2004), but fortunately optical/near-infrared follow-up
studies of known, tight T~Tauri binary systems can be used to estimate
dynamical masses for non-embedded YSOs. This was first done through
statistical means by observing small transverse motion of a number of
binary systems, and assuming random orientation of the orbits ({\it
Ghez et al.}, 1995; {\it Woitas et al.}, 2001). This led to the
conclusion that the average total system mass for T~Tauri multiple
systems is about 1.7$M_\odot$. This will be compared to dynamical
masses of embedded multiple systems in Section~3.4.

In recent years, dynamical masses were determined for individual
systems for which a substantial time coverage of their short orbital
period could be achieved, and therefore a Keplerian orbit could be
adjusted to the data ({\it Steffen et al.}, 2001; {\it Tamazian et
al.}, 2002; {\it Duch\^ene et al.}, 2003; {\it Konopacky et al.},
2006). Mass estimates range from 0.7 to 3.7 $M_\odot$ and are
generally in agreement with model predictions. Significant
uncertainties due to the limited orbital coverage and poor distance
estimates are still left, but within a few years, substantial progress
could be achieved. When this is done, it will be possible to
discriminate between evolutionary models. For more details, we refer
the reader to the review by {\it Mathieu et al.} in this volume.

\bigskip
\noindent
\textbf{ 2.4 Spectroscopic T~Tauri Binaries}
\bigskip

Due to the distance to most star-forming regions, visual binaries are
usually detected if their separation exceeds 5--10~AU. However, for
solar-type main sequence field stars, roughly a third of all
companions have tighter separations ({\it Duquennoy and Mayor}, 1991),
and cannot be spatially resolved even with current high-angular
resolution devices. These tight systems are particularly interesting,
as their formation mechanism may differ dramatically from that of
wider systems: indeed, fragmentation of prestellar cores occurs on
much larger linear scales than the sub-AU separation of spectroscopic
binaries. On the other hand, it appears that neither fission ({\it
Tohline}, 2002) nor the orbital decay of wider pairs induced by
accretion or dynamical interaction within unstable multiple systems
({\it Bate et al.}, 2002) is able to produce the observed large
population of systems with orbital periods of a few days to a few
months. One must therefore try to detect spectroscopic binaries among
T~Tauri stars, which is no easy task when considering their strong
activity, including the frequent veiling and additional line emission
induced by accretion onto, and magnetic activity in the vicinity of,
the central object.

\begin{figure}[htb]
\epsscale{0.8} \plotone{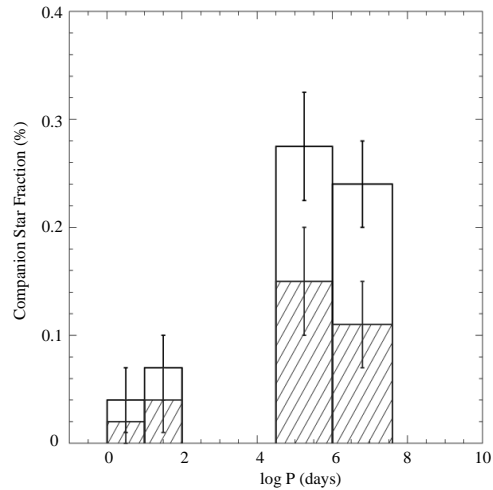}
\caption{\small Orbital period distribution for T~Tauri stars in
  several southern hemisphere star-forming regions (open histogram)
  compared to that of solar-type field stars (hatched histogram, from
  {\it Duquennoy and Mayor}, 1991); figure from {\it Melo} (2003). Note
  the non-significant excess of spectroscopic binaries.}
\end{figure}

First estimates of the frequency of T~Tauri spectroscopic binaries
were available since {\it Mathieu} (1994), but larger samples of
objects have been spectroscopically monitored since then. Most
noticeably, {\it Melo} (2003) has surveyed 59 T~Tauri stars in four
nearby star-forming regions during three campaigns over 2 years. He
found 4 new double-lined spectroscopic binaries but could not
determine their orbits due to limited time coverage. Within the
interval $1^d\leq P_{orb}\leq 100^d$, the proportion of companions is
on the order of a few percent. Once incompleteness corrections are taken
into account, {\it Melo} finds almost twice as many companions as
among field stars, however statistical uncertainties are large enough
that this is not significant (see Fig.~1). While there could be as
strong a multiplicity excess among short-period spectroscopic systems
as there is for visual companions, it will not be possible to
demonstrate it before much larger samples are monitored.

\bigskip
\noindent
\textbf{ 2.5 High-Order T~Tauri Multiple Systems}
\bigskip

As soon as the first multiplicity surveys were conducted among T~Tauri
stars, a few high-order multiple systems were identified, all triples
and quadruples. T~Tau, the eponymous low-mass pre-main sequence star,
is itself a triple system ({\it Koresko}, 2000). The number of
high-order multiple systems was too limited to pursue any statistical
analysis of their frequency and properties at the time of the review
by {\it Mathieu et al.} (2000). Among field solar-type stars, the
frequency of high-order multiples is currently being revised from 10
times to 4 times lower than that of binary systems as high-angular
resolution techniques and large surveys begin to expose the closer and
wider companions ({\it Duquennoy and Mayor}, 1991; {\it Tokovinin and
Smekhov}, 2002). Overall, high-order multiple systems may appear to be
of limited numerical strength, but their dynamical importance may be
much higher. Estimating the frequency of these high-order systems may
therefore turn out to be a more stringent constraint on fragmentation
models than the average number of companions per star, irrespective of
the system's number of components.

A first survey dedicated to the search for triple systems was
conducted by {\it Koresko} (2002) in Ophiuchus, where he focused on
already known binaries. Among 14 targets, he found 2 clear cases of
triple systems and 5 more may also be triples, suggesting a high
frequency of high-order multiples. More recently, {\it Correia et al.}
(in prep.) targeted 55 known binaries (from the list of {\it Reipurth
  and Zinnecker}, 1993) located in various star-forming regions. They
identified 15 triple and quadruple systems, i.e. a ratio of high-order
multiples to binaries on order 4, similar to recent findings of {\it
  Tokovinin and Smekhov} (2002) for main sequence systems. However, it
must be emphasized that the imaging surveys of {\it Koresko} and {\it
  Correia et al.} are only sensitive to companions wider than
$\sim10$ AU, so that the actual number of triple and quadruple systems
may be much higher. Interestingly, {\it Melo} (2003) suggested that
short-period, spectroscopic T~Tauri systems have a tendency to host
more visual companions that single stars. This result was recently
confirmed by {\it Sterzik et al.}  (2005), suggesting that the
formation of sub-AU spectroscopic systems may be related to the
presence of a third component on a stable outer orbit. For instance,
{\it Kiseleva et al.}  (1998) suggested that the combination of Kozai
cycles and tidal friction within triple systems with high relative
inclination could result in the shrinkage of the inner orbit down to
periods of only a few days. While the actual frequency of high-order
multiples among T~Tauri stars is not yet firmly established, it is
likely to place stringent constraints on star formation models.

Among the important properties of triple and higher-order multiple
systems, the relative orientation of the inner and outer orbits can
play an important role in the dynamical evolution of the systems, and
may also provide insight on their formation mechanism. Among field
triple systems, {\it Sterzik \& Tokovinin} (2002) have found a
``moderate'' alignment of the inner and outer orbits' angular momentum
vectors. Unfortunately, there are almost no T~Tauri multiple systems
for which a similar study can be performed at this point, mostly
because of the long orbital periods associated with visual binaries at
the distance of the closest star-forming regions. In the unique case
of the young hierarchical system V~773~Tau, however, {\it Duch\^ene et
al.} (2003) have argued that the inner 51~day orbital period is almost
coplanar with the outer 46~yr orbital period, although the existing
dataset is insufficient to solve all ambiguities in the orbital
solution. In coming years, the increasing number of astrometric
orbital solutions for binary young stellar objects, along with the
capacity of long-baseline interferometers to spatially resolve known
spectroscopic systems, will enable the study of the relative
inclination of orbits within pre-main sequence triple systems.
\bigskip

\noindent{\textbf{ 3. EMBEDDED MULTIPLE PROTOSTARS}}
\bigskip

As summarized in the previous section, a high multiplicity rate is
already established 1~Myr into the evolution of low-mass stars, as
demonstrated by the many observations of populations of T~Tauri stars.
However, with these observations only, the observed dichotomy between
young clusters and loose associations cannot be unambiguously
explained with a single mechanism: different pre-collapse conditions
and/or a differential dynamical evolution could be involved. Numerical
analysis has shown that close encounters within dense clusters can
substantially decrease the frequency of wide companions in less than
1~Myr (e.g., {\it Kroupa}, 1995), and that non-hierarchical systems
decay to stable configurations through few-body interactions in less
than a hundred crossing times, i.e., in 0.1~Myr or even less (e.g.,
{\it Anosova}, 1986; {\it Sterzik \& Durisen}, 1998). It is therefore
critical to conduct multiplicity studies in the youngest possible
stellar populations in order to determine the ``initial conditions''
of the evolution of multiple systems. This is why the observational
effort in this field has shifted in recent years towards the study of
the multiplicity of even younger systems, namely embedded protostellar
objects. The existence of extremely young (Class~0) multiple systems
(e.g., {\it Wootten}, 1989; {\it Loinard}, 2002; {\it Chandler et
al.}, 2005) shows that the formation of multiples most probably occurs
very shortly after the apparition of the initial protostellar seeds.
Class~0 and Class~I sources represent objects whose age is believed to
be on the order of a few $\times$ $10^4$~yr and a few $\times$
$10^5$~yr, respectively.  While they may already be too old to be
considered pristine from the point of view of dynamical evolution,
these objects provide an opportunity to probe an intermediate stage of
the star formation process, where some evolution has already taken
place but, hopefully, is not yet over. Furthermore, it is possible
that the youngest (Class~0) protostars have suffered only very little
evolution. In this section, we focus on such embedded multiple systems
in order to assert some of their basic properties, and how they
compare to more evolved T~Tauri multiple systems.

\bigskip
\noindent
\textbf{ 3.1 A Combination Of Observational Approaches}
\bigskip

Already at the time of {\it Protostars and Planets IV}, a few embedded
multiple systems were known (see, e.g., the discussion of L1551~IRS5
in {\it Mathieu et al.}, 2000). However, the first statistical surveys
of such objects have only been conducted in the last few years with
the advent of a newer generation of instruments. Because they are
still enshrouded in their dusty cocoons, the youngest protostars are
not detectable at visible wavelengths. They are often dim even in the
near- and mid-infrared, and emit most of their luminosity at
far-infrared and sub-millimeter wavelengths, where the angular
resolution currently available remains limited. Fortunately, high
angular resolution techniques are now available in the near-infrared,
mid-infrared, and radio domains.

The most embedded, Class~0 protostars are often associated with
relatively bright and compact radio emission. Indeed, they frequently
power supersonic jets that generate free--free emission detectable at
radio wavelengths ({\it Rodr\'{\i}guez}, 1997), they are surrounded by
accretion disks whose thermal dust emission is sometimes still
detectable in the centimeter regime ({\it Loinard et al.}, 2002; {\it
Rodr\'{\i}guez et al.}, 1998, 2005a), and they often have active
magnetospheres ({\it Dulk}, 1985; {\it Feigelson and Montmerle}, 1999;
{\it Berger et al.}, 2001; {\it G\"udel}, 2002). The former two
mechanisms produce emission on linear scales of tens to hundreds of
astronomical units, whereas the non-thermal emission related to active
magnetospheres is usually thought to result from the interaction of
mildly relativistic electrons with the strong magnetic fields (a few
kGauss) which are often present at the surface of young, low-mass
stars (e.g., {\it Valenti and Johns-Krull}, 2004; {\it Symington et
al.}, 2005). This process, therefore, produces emission on very small
scales, typically a few stellar radii. Interferometric radio
observations (7~mm $\lesssim\lambda\lesssim6$~cm) can supply images of
YSOs with such high angular resolution and excellent astrometric
quality: NRAO's {\em Very Large Array} (VLA) and {\em Very Long
Baseline Array} (VLBA) connected interferometers provide typical
angular resolution of 0\farcs1 and 0\farcs001 in combination with
astrometric accuracies of 0\farcs01 and 0\farcs0001, respectively. The
combination of these two assets makes it possible with radio
interferometry to identify tight binaries among those protostars which
emit centimeter radio waves, and study their orbital motions.

Class~I protostars, which are more evolved and less deeply embedded,
are strong mid-infrared emitters. The 10~$\mu$m radiation from each
component does not originate from the star itself, but from a
``photosphere" of surrounding dust heated to several hundred degrees.
According to the radiative transfer model of {\it Chick and Cassen}
(1997), the 10~$\mu$m photosphere is located about 1~AU from a
low-mass protostar. In the mid-infrared regime, direct imaging on the
newest generation of instruments on 6-10m telescopes provides deep,
diffraction-limited ($\lesssim$0\farcs3) images that are extremely
sensitive to protostars, and whose spatial resolution largely surpasses
current space capabilities. At such a spatial resolution, the
individual components of Class~I sources should be point sources in
the mid-infrared, and easy to disentangle from one another.

As far as ground-based observations are concerned, the near-infrared
regime currently offers the best combination of spatial resolution
(diffraction limit on the order of 0\farcs05 for 8-10m telescopes),
sensitivity, and field of view. The spectral energy distribution (SED)
of Class~I protostars extends into the near-infrared and, as in the
mid-infrared, the light emitted by these objects comes from a very
small photosphere that remains unresolved. Near-infrared observations
have therefore become one of the most powerful approaches to study the
multiplicity of Class~I protostars. Observations in this regime not
only allow the discovery of very tight companions, but also provide a
probe of the evolutionary state of individual components: the latter
is directly related to the spectral index of YSOs in the near- to
mid-infrared regime ({\it Lada}, 1987).

Taking advantage of these complementary techniques, we will now
discuss several recent results concerning young embedded multiple
systems: their average multiplicity rate, their evolutionary status,
and high-precision astrometric follow-up studies (orbital motions,
possible dynamical decays).

\bigskip
\noindent
\textbf{ 3.2 The Multiplicity Of Low-Mass Embedded Protostars}
\bigskip

\noindent{\it 3.2.1. Radio Imaging Surveys}. One of the first
systematic surveys for multiplicity of embedded YSOs was conducted by
{\it Looney et al.} (2000) with the BIMA interferometer at
2.7~mm. Although companions as wide as 15000~AU could be identified in
this survey, one must be cautious that such wide systems may not be
physically bound. Considering a 2000~AU upper limit for projected
separation, a value frequently used for other multiplicity surveys,
and focusing on the Class~0 and Class~I sources in their sample, 3 out
of the 16 independent targets actually are binaries and none is of
higher multiplicity. Still, this is quite a high multiplicity rate
considering the relatively large lower limit on projected separation
(60-140~AU depending on the molecular cloud): it is somewhat higher
than the rate observed for solar-like field stars. Furthermore, it
must be emphasized that millimeter observations have a limited
sensitivity to YSOs in multiple systems with separations of a few
hundred AUs because their disks are much reduced as a result of
internal dynamics ({\it Osterloh and Beckwith}, 1995; {\it Jensen et
al.}, 1994). Overall, the limited size of the sample studied by {\it
Looney et al.} (2000) warrants statistically robust conclusions, and
the main outcome of this pioneering work was to confirm the prevalence
of multiple systems on a wide range of spatial scales among the
youngest embedded protostars.

In parallel to this study, {\it Reipurth} (2000) also emphasized that
low-mass embedded objects are frequently binary or multiple. In his
study of 14 sources driving giant Herbig-Haro flows (mostly Class~0
and Class~I sources), based on a variety of high-angular resolution
datasets, he concluded that they had an observed multiplicity rate
between 80 and 90\%, and that more than 50\% of them were higher order
multiples. He further suggested that strong outflows could be the
consequence of accretion outburst occurring during the dynamical decay
of unstable high-order multiple systems. If this statement is correct,
the extremely high multiplicity rate may be overestimated due to a
selection bias. In posterior VLA continuum studies on a larger sample
of mostly Class~I sources and with a uniform observing strategy, {\it
Reipurth et al.} (2002, 2004) found a binary frequency of 33\% in the
separation range from 0.5$''$ to 12$''$ for a sample of 21 embedded
objects located between 140 and 800pc from the Sun. Within the
uncertainties, this binary frequency is comparable to the observed
binary frequency among T~Tauri stars in a similar separation
range. Among this new sample, 4 out of 7 objects that drive giant
molecular flows were found to have companions, a marginally higher
rate that is not sufficient for any conclusions to be reached on the
multiplicity-outflow connection.
\bigskip

\noindent{\it 3.2.2. Near-Infrared Imaging Surveys}. Independent
systematic surveys of the multiplicity of embedded protostars were
later conducted at near-infrared wavelengths (1--4~$\mu$m) on samples
of several tens of Class~I and ``flat spectrum'' embedded sources. A
first series of surveys ({\it Haisch et al.}, 2002, 2004; {\it
Duch\^ene et al.}, 2004) were conducted using wide-field near-infrared
cameras, which permitted seeing-limited observations. Sources in the
Perseus, Taurus-Auriga, Chamaeleon, Serpens and $\rho$ Ophiuchi
molecular clouds were targeted. To derive a robust multiplicity rate,
we define here a ``restricted'' companion star fraction by focusing on
the 300--1400~AU projected separation range, for which all targets
have been observed, and by retaining only those companions which
satisfy $\Delta K \leq 4$~magnitude, following {\it Haisch et al.}
(2004). Merging all surveys into one large sample, there are 119
targets, for which 19 companions are identified within these
limits. This corresponds to a multiplicity rate of 16.0$\pm$3.4\%,
with all clouds presenting entirely consistent rates. The multiplicity
rate found for Class~I sources in Taurus and $\rho$ Ophiuchi is in
excellent agreement with those obtained for T~Tauri stars in the same
star-forming regions, and is about twice as high as that observed for
late-type field dwarfs (see Fig.~2). A few systems present very large
near-infrared flux ratios (up to $\Delta K \sim 6$~magnitudes); if the
physical nature of these pairs is confirmed, these companions could be
candidate proto-brown dwarfs, whose frequency should then be compared
to the (very rare) occurrence of wide star-brown dwarf systems among
field stars.  Alternatively, these systems could be understood as the
association of two objects whose evolutionary states are different,
with the fainter star still being much more deeply embedded. Only
subsequent mid-infrared imaging and/or high resolution spectroscopy
will help disentangle these two options.

\begin{figure}[htb]
\epsscale{0.9} \plotone{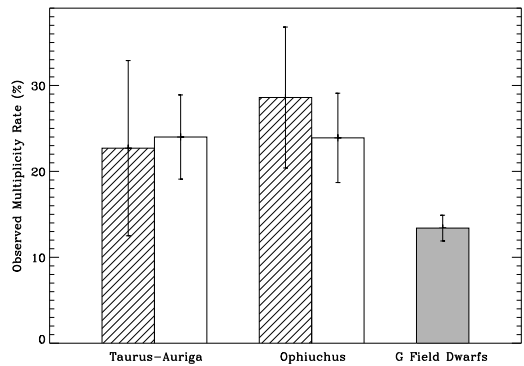}
\caption{\small Observed multiplicity rates in the projected
  separation range 110--1400~AU for Class~I protostars (hatched
  histograms) and T~Tauri stars (open histogram) in the Taurus-Auriga
  and Ophiuchus molecular clouds; adapted from {\it Duch\^ene et al.}
  (2004). The multiplicity rate for field solar-type stars from {\it
  Duquennoy and Mayor} (1991) is shown for reference.}
\end{figure}

To study in more detail the frequency and properties of multiple
systems, higher spatial resolution observations are required. Indeed,
such observations enable the discovery of many more companions and, in
particular, of a number of stable hierarchical triple and higher-order
multiple systems. {\it Duch\^ene et al.} (in prep.) have recently
conducted an adaptive optics imaging survey of 44 Class~I protostars
located in the Taurus-Auriga, $\rho$ Ophiuchi, Serpens, and Orion
(L~1641) molecular clouds. The diffraction-limited images they
obtained on the 8m-VLT allowed companions as close as 0\farcs1
($<20$~AU in the closest clouds) to be resolved, providing an order of
magnitude improvement in spatial resolution from previous surveys, and
identifying a dozen subarcsecond companions that were not known
previously. Combining these observations with direct images of the
same sources from previous surveys, and concentrating on the
36--1400~AU separation range, there are a total of 23 companions
fulfilling the $\Delta K \leq 4$~magnitude criteria, representing a
total multiplicity rate of 52.2$\pm$7.5\%. Within current statistical
uncertainties, observations are consistent with the hypothesis that
Class~I multiple systems have essentially the same properties in all
clouds. Among this high-angular resolution sample, 6 systems are
triple, including 5 in which the ratio of projected separations is
high enough to consider that they are hierarchically stable. No
higher-order system was identified in this survey. The observed
proportion of triple systems ($\sim14\%$) is comparable to that
observed among T~Tauri stars (see Section~2.5), and may be higher than
the proportion of such systems among field stars.
\bigskip

\noindent{\it 3.2.3. Spectroscopic Surveys}. The radio and
near-infrared surveys conducted so far have been able to identify
companions with separations of several tens to hundreds of
AUs. However, as mentioned already, many companions to low-mass field
stars are on much tighter orbits. While VLBA observations already
provide an opportunity to resolve some tight spectroscopic binary
systems if they are strong radio emitters (see Section~3.4), it
remains difficult to conduct a systematic analysis of the frequency of
spectroscopic binaries among embedded protostars with imaging
techniques. An alternative approach is to obtain high resolution
spectra of protostars to identify spectroscopic binaries.  Despite the
difficulty induced by the partially opaque envelope that surrounds
these objects, this spectroscopic effort is now ongoing, both at
visible and near-infrared wavelengths (see the chapter by {\it Greene
et al.}  in this volume). Using single-epoch radial velocity
measurements for a sample of 31 Class~I and flat-spectrum protostars
in several star-forming regions, {\it Covey et al.}  (2006) found 4
objects whose radial velocities significantly depart from that of the
surrounding local gas velocity. None of the objects were found to be
double-line spectroscopic binaries, and their discrepant radial
velocities either indicate that they have been ejected after an
unstable multi-body interaction, or that they are single-line
spectroscopic binaries. Long-term monitoring of these objects will
reveal their true nature. At any rate, this preliminary study shows
that systematic searches for spectroscopic binaries among embedded
protostars are now feasible. We can therefore hope to rapidly
complement existing imaging surveys to almost completely cover the
entire range of orbital periods from a few days up to separations of
several thousand AU. Comparing such surveys to the known properties of
field stars would be important to determine whether the tightest
systems actually form through a different mechanism than wide pairs.
\bigskip

\noindent{\it 3.2.4. Implications For Star Formation
Scenarios}. Overall, these surveys for multiplicity among embedded
protostars consistently support the general scenario in which all
star-forming regions produce a very high fraction of binary and
higher-order multiple systems, at least as high as that observed among
T~Tauri stars in the most binary-rich regions like
Taurus-Auriga. While there is marginal evidence for a decrease in
multiplicity rate between the most and least embedded protostellar
sources ({\it Duch\^ene et al.}, 2004), no significant evolution of
the multiple system population has been found in regions like Taurus
or Ophiuchus within the $\lesssim1$~Myr timescale during which
protostars evolve into optically bright T~Tauri stars. The absence of
mini-clusters of 5 or more sources within $\sim2000$~AU implies that
if cores frequently fragment into many independent seeds, they must
decay into unbound stable configurations within a very short
timescale, on the order of $\sim10^5$~yrs at most. The relatively low
number of protostars found to be single in the surveys presented here
($\lesssim50$\%), however, suggests that cores can rarely result in
the formation of unstable quadruple or higher order multiples,
supporting the point of view recently presented by {\it Goodwin and
Kroupa} (2005). This and other star formation models are further
discussed in Section~4.2.

One of the most intriguing results that arises from these surveys is
the finding that there seems to be no influence of environmental
conditions on the multiplicity of embedded protostars, as opposed to
what is observed for T~Tauri multiple systems. Namely, Class~I
protostars in Orion (in the L1641 cloud) show as high a multiplicity
excess over field stars as all categories of YSOs in the Taurus and
$\rho$ Ophiuchi clouds. This seems to favor a scenario in which the
end result of core fragmentation is independent of large scale
physical conditions, but rather is sensitive only to small-scale
physics which may well be very similar in all molecular clouds. As
already discussed by {\it Kroupa et al.} (1999), dynamical disruptions
among an initial population of binary systems can account for the
deficit of wide binaries observed for optically bright YSOs in dense
clusters, even though the population of ``primordial multiple
systems'' has universal properties. Based on the present observations,
we may conclude that such disruptive encounters, which represent a
distinct process from the internal decay of unstable multiple systems,
is likely to occur in the densest clusters on a timescale of $\sim$
few $\times$ $\sim10^5$~yrs.
\bigskip

\noindent
\textbf{ 3.3 Evolutionary Status Within Multiple Protostars}
\bigskip

{\it Haisch et al.} (2006) have recently obtained new mid-infrared
observations of 64 Class~I and flat-spectrum objects in the Perseus,
Taurus, Chamaeleon~I and II, $\rho$ Ophiuchi, and Serpens dark
clouds. They detected 45/48 (94\%) of the single sources, 16/16
(100\%) of the primary components, and 12/16 (75\%) of the
secondary/triple components of the binary/multiple objects
surveyed. The 10~$\mu$m fluxes, in conjunction with $JHKL$ photometry
from {\it Haisch et al.}  (2002, 2004), were used to construct SEDs
for the individual binary/multiple components. Each source was
classified using the least squares fit to the slope of its SED between
2.2 and 10~$\mu$m in order to quantify their nature. The
classification scheme of {\it Greene et al.} (1994) has been adopted
in our analysis as it is believed to correspond well to the physical
stages of evolution of YSOs ({\it e.g. Andr\'{e} and Montmerle},
1994). A Class~I object is one in which the central YSO has attained
essentially its entire initial main-sequence mass, but is still
surrounded by a remnant infalling envelope and an accretion
disk. Flat-spectrum YSOs are characterized by spectra that are
strongly veiled by continuum emission from hot, circumstellar
dust. Class~II sources are surrounded by accretion disks, while
Class~III YSOs have remnant, or absent, accretion disks. Thus, the
progression from the very red Class~I YSO {\bf $\rightarrow$} flat
spectrum {\bf $\rightarrow$} Class~II {\bf $\rightarrow$} Class~III
has been frequently interpreted as representing an evolutionary
sequence, even though {\it Reipurth} (2000) has suggested that more
violent transitions from the embedded to the optically-bright stages
could occur when components are ejected from unstable multiple
systems.  \bigskip

\noindent{\it 3.3.1. Nature Of ``Mixed" Systems}. While the composite
SEDs for all YSOs in the {\it Haisch et al.}  (2006) study are either
Class~I or flat-spectrum, the individual source components sometimes
display Class~II, or in one case Class~III, spectral indices. The SED
classes of the primary and secondary components are frequently
different. For example, a Class~I object may be found to be paired
with a flat-spectrum source, or a flat-spectrum source paired with a
Class~II YSO. Such behavior is not consistent with what one typically
finds for T~Tauri stars, where the companion of a classical T~Tauri
star also tends to be a classical T~Tauri star ({\it Prato and Simon},
1997; {\it Duch\^ene et al.}, 1999), although mixed pairings have been
previously observed among Class~II YSOs (e.g., {\it Ressler and
  Barsony} 2001). Indeed, there appears to be a higher proportion of
mixed Class I/Flat-Spectrum systems (67\%, {\it Haisch et al.}, 2006)
than of mixed CTTS/WTTS systems (25\%, {\it Hartigan and Kenyon},
2003; {\it Prato et al.}, 2003; {\it McCabe et al.}, 2006; see review
by {\it Monin et al.} in this volume).

While several of the Class I/Flat-Spectrum binary components lie in
regions of a $JHKL$ color-color diagram which are not consistent with
their SED classes, they all lie in their expected locations in a $KLN$
color-color diagram. In fact, in this diagram, one can see a clear
progression from the very red Class I YSO {\bf $\rightarrow$} flat
spectrum {\bf $\rightarrow$} Class II ({\it Haisch et al.}, 2006).
Taken together with the above discussion, this demonstrates the fact
that while in most cases the SED class reflects the evolutionary state
of the YSO, there may be instances in which the SED class does not
yield the correct evolutionary state. The rigorously correct way to
determine an objects' evolutionary state is to obtain multi-wavelength
imaging data for each source and quantitatively compare these data to
models produced using 3D radiative transfer codes ({\it Whitney et
al.}, 2003).

Visual extinctions, A$_{v}$, have been determined for all
binary/multiple components, except the Class~I sources, for which
accurate dereddened colors cannot be derived using infrared
color-color diagrams. In general, the individual binary/multiple
components suffer very similar extinctions, A$_{v}$, suggesting that
most of the line-of-sight material is either foreground to the
molecular cloud or circumbinary in nature.
\bigskip

\noindent{\it 3.3.2. Notes On Selected Objects}. Among the various
mid-infrared surveys to date, several sources deserve specific
mention. A detailed study of WL~20, the only non-hierarchical triple
system among embedded protostars, by {\it Ressler and Barsony} (2001)
and {\it Barsony et al.} (2002) has suggested that disk interaction
has resulted in enhanced accretion onto one component of this system,
WL~20S. This tidally-induced disk disturbance could explain the
Class~I SED of this object, although it is probably coeval, at an age
of several million years, with its Class~II companions. On the other
hand, the recent high-angular resolution images obtained in the
near-infrared by {\it Duch\^ene et al.} (in prep) as part of their
multiplicity survey revealed a completely unexpected morphology that
does not seem to be consistent with a $\sim1$ AU opaque envelope
photosphere. Rather, it seems like WL~20~S is a normal T~Tauri system
whose circumstellar material has such a geometry that only scattered
light reaches the observer shortwards of $\sim5\mu$m, emphasizing the
need for high angular resolution images to determine the actual nature
of an embedded YSO.

IRS~43 (also known as YLW~15) in Ophiuchus was found to be a binary
VLA source with 0\farcs6 separation ({\it Girart et al.}, 2000).
IRS~43 is also part of a wide-binary system with GY~263 ({\it Allen et
  al.}, 2002). {\it Haisch et al.} (2002) find IRS~43 to be multiple at
10~$\mu$m but single in the near-infrared. The brighter mid-infrared
source in IRS~43 corresponds to VLA~2 and the heavily-veiled, Class~I
near-infrared source. VLA~1 is an embedded protostar, undetected in
the near-infrared, and possibly in the Class~0 to Class~I transition
and powering a Herbig-Haro outflow. Its mid-infrared emission appears
slightly resolved with a diameter of $\sim$16~AU, possibly tracing
circumstellar material from both the envelope and the disk ({\it
  Girart et al.}, 2004). Both VLA/mid-infrared sources associated with
IRS~43 are embedded in extended, faint near-infrared nebulosity imaged
with HST/NICMOS ({\it Allen et al.}, 2002). Strikingly, the near- to
mid-infrared properties of YLW~15 suggest that VLA~1 is a more
embedded YSO, or alternatively, less luminous than VLA~2, whereas
orbital proper motions of this binary system by {\it Curiel et al.}
(2003) indicate that VLA~1 is more massive than VLA~2. This is
apparently against the expected evolutionary scenario, in which one
expects that the more massive YSO in a binary system is the more
evolved and more luminous YSO.

Another source, ISO-Cha~I~97 in Chamaeleon~I, was detected as a single
star in the near-infrared; however, mid-infrared observations have
revealed that this source is in fact binary ({\it Haisch et al.},
2006). The $K$-band sensitivity limit from {\em Haisch et al.} (2004),
combined with its 10~$\mu$m flux, yields an extremely steep lower
limit to the spectral index that places ISO-Cha~I~97 in a class of YSO
that has heretofore been rarely known.  Three such objects have been
recently reported, the Class~0 object Cep~E~mm ({\it Noriega-Crespo et
al.}, 2004), source~X$_{E}$ in R~CrA ({\it Hamaguchi et al.}, 2005),
and source L1448~IRS~3~A ({\it Ciardi et al.}, 2003; {\it Tsujimoto et
al.}, 2005). Further very steep spectrum YSOs are expected to be
discovered with the Spitzer Space Telescope.

Spatially resolved mid-infrared spectroscopy of the
Class~I/flat-spectrum protostellar binary system, SVS~20 in the
Serpens cloud core, has been recently obtained by {\it Ciardi et al.}
(2005). SVS~20~S, the more luminous of the two sources, exhibits a
mid-infrared emission spectrum peaking near 11.3~$\mu$m, while
SVS~20~N exhibits a shallow amorphous silicate absorption spectrum
with a peak optical depth of $\tau \sim0.3$. After removal of the
line-of-sight extinction by the molecular common envelope, the
``protostar-only'' spectra are found to be dominated by strong
amorphous olivine emission peaking near 10~$\mu$m. There is also
evidence for emission from crystalline forsterite and enstatite
associated with both SVS~20~S and SVS~20~N. The presence of
crystalline silicate in such a young binary system indicates that the
grain processing found in more evolved Herbig Ae/Be and T~Tauri
pre-main-sequence stars likely begins at a relatively young
evolutionary stage, while mass accretion is still ongoing. A third
component to the system was found by {\it Duch\^ene et al.} (in
prep.), making the analysis of the system even more complex.

Finally, {\it Meeus et al.} (2003) have presented mid-infrared
spectroscopy of three T Tauri stars in the young Chamaeleon I dark
cloud, CR Cha, Glass I, and VW Cha, in which the silicate emission
band at 9.7~$\mu$m is prominent. This emission was modeled with a
mixture of amorphous olivine grains of different size, crystalline
silicates, and silica. The fractional mass of these various components
was found to change widely from star to star. While the spectrum of CR
Cha is dominated by small amorphous silicates, in VW Cha (and in a
lesser degree in Glass I), there is clear evidence of a large amount
of processed dust in the form of crystalline silicates and large
amorphous grains. Interestingly, the two objects with an ``evolved''
dust population are associated with a tight companion, leading to the
intriguing speculation that multiplicity may accelerate dust
processing in cirucmstellar disks.  \bigskip

\noindent
\textbf{ 3.4 High-Accuracy Astrometry Of Embedded Multiples}
\bigskip

As pointed out in Section~3.1, VLA and VLBA observations can provide a
view of embedded protostars with an unsurpassed resolution and
astrometric accuracy. This allows a number of studies that are
impossible to conduct with shorter wavelength observations. We summarize
here some recent results that take full advantage of VLA and VLBA
capabilities to study embedded multiple systems.
\bigskip

\noindent{\it 3.4.1. Orbital Motion Within Embedded Multiple
Systems}. For a very limited subset of tight radio binaries associated
with low-luminosity Class~0 and Class~I deeply embedded sources,
orbital motions have now been detected by comparing images taken at
several epochs. It is very important to determine the system masses in
order to constrain the time evolution of the central sources, and, for
instance, to determine the fate of the material located in the
circumstellar envelopes of Class~I sources. Comparable estimates are
already available for T~Tauri stars (see Section~2.3).

Due to slightly poorer spatial resolution of VLA observations compared
to the highest angular resolution near-infrared datasets, the embedded
multiple systems for which orbital motion was detected typically have
orbital periods of hundreds of years. Therefore, the multi-epoch
observations to date only cover a small fraction of the orbit, and the
exact orbital parameters cannot be derived. However, a reasonable
estimate of the mass of the system can still be obtained assuming
circular orbits, if the inclination can be guessed from independent
means. The four sources where this could be achieved are
IRAS~16293--2422 ({\it Loinard}, 2002; {\it Chandler et al.}, 2005)
and YLW~15 ({\it Curiel et al.}, 2003) in $\rho$~Ophiuchi, and L~1527
({\it Loinard et al.}, 2002) and L~1551 ({\it Rodr\'{\i}guez et al.},
2003) in Taurus-Auriga. The mass estimates are respectively:
2.8$\pm$0.7$M_\odot$; 1.7$\pm$0.8$M_\odot$, 1.0$\pm$0.5$M_\odot$, and
1.2$\pm$0.5$M_\odot$.

The mean value is therefore 1.7 $\pm$ 0.7 $M_\odot$, confirming that
these low-luminosity, deeply embedded protostars are most likely the
precursors of solar-type stars. Noticeably, this average mass is not
significantly lower than that derived for T~Tauri binary systems.
Although small number statistics and selection biases preclude
definitive conclusions at this stage, this seems to imply that the
mass of the remnant envelopes in these systems are already
significantly lower than the stellar seeds themselves; this is
expected for Class~I sources but may be more surprising for Class~0
sources. Future astrometric follow-up studies on these and other
embedded tight binary systems will help clarify this issue.  \bigskip

\noindent{\it 3.4.2. Spectroscopic Systems Resolved With
Interferometry}. Most radio observations of embedded protostars have
been conducted with the VLA. As mentioned above, the VLBA offers an
angular resolution and an astrometric precision two orders of
magnitude better than the VLA, but can detect YSOs that emit bright
and compact non-thermal radio emission. Although all protostars
probably do generate non-thermal radiation at some level, those that
are currently easily detectable represent only a limited
sample. Recently, Loinard and coworkers have started to monitor 10
protostars in the Taurus and $\rho$~Ophiuchi star-forming regions that
are known to be non-thermal radio sources. Interestingly, at least 2
of these systems were found to be multiple with separations of only a
few times 0\farcs001, or a few tenths of an AU at the distance of
these molecular clouds. The clearest case is that of V773~Tau which
was previously known to be a spectroscopic binary ({\it Welty}, 1995),
and had been found to be a double in previous VLBI observations ({\it
Phillips et al.}, 1996).  While V773~Tau is a T~Tauri system, more
embedded systems are among the sample studied in this survey and, if
they are found to have a tight companion, they would represent the
earliest stage at which spectroscopic binaries can actually be
spatially resolved. In parallel to this VLBA approach, near- and/or
mid-infrared long baseline interferometers (such as Keck or VLTI)
could resolve these systems in the near future, allowing for powerful
infrared-to-radio analysis. In any case, these very tight systems
clearly have a much shorter period ($\lesssim$1yr) than the embedded
multiple systems resolved so far, so one should be able to accurately
measure their masses in a comparatively shorter time.  \bigskip

\noindent{\it 3.4.3. Hints Of Disruption?} With high-angular
resolution and high astrometric accuracy, following the relative
motion within multiple systems may reveal departures from bound
Keplerian orbits. In particular, the possibility of studying the
internal dynamics of unstable multiple systems, if some can be found,
is extremely appealing. A special case of this decay is the violent
disruption of a triple system into an ejected single star and a stable
binary system.  Only radio observations with the VLA combine high
resolution and astrometric accuracy with a relatively long time
baseline. It is therefore not a surprise that the first two claims of
possible disruption of young multiple systems have come from such
observations.

Using multi-epoch archival VLA observations of the triple system T
Tauri over almost 20 years, {\it Loinard et al.} (2003) have found
that the trajectory of one of its components could not be easily
explained in terms of a stable Keplerian orbit. Follow-up
near-infrared observations have cast doubt upon this conclusion as the
orbit of the putative ejected component has been seen to slow down and
curve as if it were on a Keplerian orbit ({\it Furlan et al.}, 2003;
{\it Tamazian}, 2004; {\it Beck et al.}, 2004). However, no satisfying
orbit has been fit simultaneously to all infrared and radio
observations, and the exact correspondence of the detected sources at
both wavelengths is still under debate ({\it Johnston et al.}, 2004,
2005; {\it Loinard et al.}, 2005). Following the system in both
wavelength regimes for a few more years should yield a final
conclusion to this issue.

\begin{figure}[htb]
\epsscale{0.8} \plotone{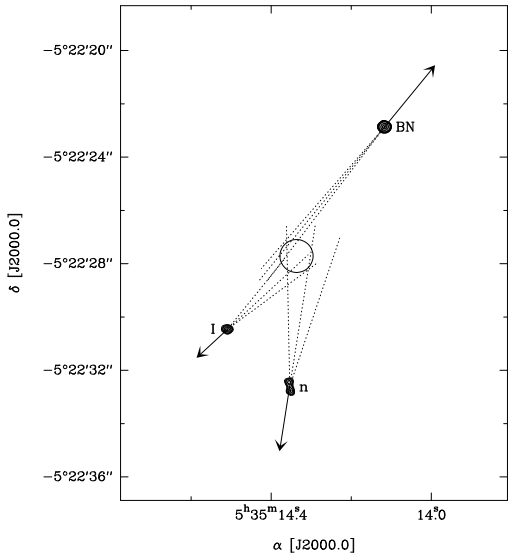}
\caption{\small Proper motion of three radio sources in the
  Becklin-Neugebauer/Kleinman-Low region, from {\it G\'omez et al.}
  (2006). They all trace back to the same point in space and time,
  500~years ago.}
\end{figure}

More recently, {\it Rodr\'{\i}guez et al.} (2005b) and {\it G\'omez et
  al.} (2006) have shown that three of the four compact radio sources
around the Orion Becklin-Neugebauer/Kleinmann-Low region are moving
away from a common point of origin, where they must have been located
about 500 years ago (Fig.~3). These three radio sources are apparently
associated with relatively massive young stars (M $>$ 8 $M_\odot$),
suggesting that a massive multiple system disintegrated around that
time. Although a different point of origin for the Becklin-Neugebauer
object and an eightfold longer timescale has been advocated by {\it
  Tan} (2004), there is little doubt that this system was formed as an
unstable multiple system, and has very recently experienced a dynamical
ejection event.

These two cases remain open to discussion, and are insufficient to
assess the exact relevance of few-body encounters to the final rates
of multiplicity in stellar systems. However, they demonstrate the
potential of radio interferometry to tackle this important
issue, and should be pursued in upcoming years. \bigskip

\noindent{\textbf{ 4. CORE FRAGMENTATION AND EARLY DYNAMICAL
    EVOLUTION OF MULTIPLE SYSTEMS}} 
\bigskip

In this section, we review some of the most recent numerical
simulations that aim at following the processes of collapse and
fragmentation of a prestellar core as well as the subsequent dynamical
interactions within the multiple systems resulting from
fragmentation. More specifically, we concentrate on three different
models of star formation, namely those by Goodwin and collaborators
({\it Goodwin et al.}, 2004a,b), Sterzik and collaborators ({\it
Sterzik and Durisen}, 1995, 1998; {\it Durisen et al.}, 2001; {\it
Sterzik et al.}, 2003), and Delgado-Donate and collaborators ({\it
Delgado-Donate et al.}, 2004a,b), with special emphasis on the latter
two. A more detailed description of other simulations can be found in
two other chapters in this volume ({\it Whitworth et al.} and {\it Goodwin
et al.}).

\bigskip
\noindent
\textbf{ 4.1 A Brief Overview Of Current Simulations}
\bigskip

For some time, numerical models with predictive power on the
statistical properties of young stars (e.g. multiplicity fractions,
mass ratio, semi-major axis distributions) had to rely on pure N-body
integration of the break-up of small clusters of point masses ({\it
Sterzik and Durisen}, 1995, 1998; {\it Durisen et al.}, 2001). The
masses, location, and velocities of the stars had to be selected at the
outset, and subsequently the orbital evolution was calculated. This
approach to multiple star `formation' has the advantage of being easy
and fast to calculate, so that many realizations of the same initial
conditions could be run. However, it completely neglects the modeling
of gas fragmentation, collapse, and accretion, a highly demanding task
from a computational point of view. Yet, gas is a fundamental
ingredient of the star formation process, not only during the
fragmentation and collapse stage, but also during the embedded phase
of the life of a star. Gaseous material accumulates in the form of
accretion disks around the protostars, and these disks can modify
substantially both the orbital parameters of a protobinary ({\it
Artymowicz and Lubow}, 1996; {\it Bate and Bonnell}, 1997; {\it Ochi et
al.}, 2005) and the outcome of dynamical encounters with other cluster
members ({\it McDonald and Clarke}, 1995). Furthermore, under adequate
physical conditions, disks can fragment, and in doing so, produce a
second generation of objects ({\it Gammie}, 2001; {\it Lodato and Rice},
2005). Gas also acts on the large scale throughout the embedded phase
of star formation by providing a substantial contribution to the
gravitational potential of the system. In this manner, gas can affect
the mass evolution and motion of both single and multiple stars, hence
the binary pairing outcomes, through the action of gravitational drag
({\it Bonnell et al.}, 1997, 2003; {\it Delgado-Donate et al.}, 2003).
Although they prioritize the N-body dynamics over the gas dynamical
processes, the simulations by Stezik and collaborators provide useful
constraints on what effects dynamical interactions alone can have on
the star formation process. Interestingly, these calculations provide
the best match to date to the mass dependence of the multiplicity
fraction of stars, as is constrained by our present observational
knowledge. They do so by means of a 2-step procedure ({\it Durisen et
al.}, 2001), whereby the stellar masses are picked randomly from a
stellar mass function, subject to the additional constraint that the
total cluster mass equals a value also picked randomly from a cluster
mass function. This way, they alleviate the usually steepening effect
of the process known as dynamical biasing -- i.e. the strong trend of
the two most massive stars in a cluster to pair together -- on the
multiplicity-vs-primary mass curve, by having a significant number of
clusters where the two most massive stars have both low masses. This
major success of N-body models is unmatched by current gas dynamical
calculations, which so far are just able to give a positive but too
steep dependence of the multiplicity fraction on primary mass (see
Section~4.2). This success should not mask an obvious caveat, however:
it remains unclear at present how a cluster may break into subunits
which fragment into stars with the appropriate mass spectrum set by
the 2-step process.

Early star formation models that included the effect of gas did so to
study the formation of binary stars from clouds subject to some kind
of specific initial instability (see the reviews by {\it Sigalotti and
Klapp}, 2001 and {\it Tohline}, 2002). These models have been of great
importance, but a caveat remained: they produce a low number of
objects in a more or less predictable fashion. Other models tried to
take into account large numbers of stars embedded in a big gas cloud
({\it Bonnell et al.}, 2001), e.g. by utilizing point masses with the
ability to accrete and interact with the gas and other stars (`sink
particles'; {\it Bate et al.}, 1995), but once more, with positions and
velocities selected at the outset. These models focused mostly on the
study of the resulting initial mass function. The purely gas dynamical
models had to be refined and taken to a larger scale, while the aim of
point-masses-in-gas models had to shift to the study of the properties
of multiple stars, if star formation models were to match the
predictive power of N-body models. The earliest model to take such
step was that by {\it Bate et al.}  (2003), who applied more general
`turbulent' initial conditions to a relatively large (for theory
standards) $50M_\odot$ cloud and followed its fragmentation and
collapse down to the opacity limit for fragmentation. For higher
densities, pressure-supported objects were replaced by `sink
particles' and, thus, the simulation could be followed well beyond the
formation of the first objects. This calculation showed the power of
the combination of more realistic initial conditions and a refined
numerical scheme blending gas with N-body dynamics and, beyond any
doubt, it meant a great leap forward in star formation studies; but,
obviously, it had some shortcomings too. Among them was the high
computational expense involved, and the fact that the evolution of the
cloud could not be followed for as long as it would be desirable in
order to ensure the stability of most multiples. Thus, complementary
calculations were necessary, and these were performed mainly by
Delgado-Donate and collaborators and Goodwin and collaborators. The
{\it Goodwin et al.} (2004a,b) and {\it Delgado-Donate et al.}
(2004a,b) simulations model the fragmentation of small
($\approx5M_\odot$) molecular clouds subject to different degrees of
internal turbulent motion. Their models basically differ in the
resolution employed, and the different subsets of parameter space
studied: most importantly for this review, {\it Goodwin et al.} study
subsonic turbulence whereas {\it Delgado-Donate et al.} impose
supersonic to hypersonic random velocity fields. Both sets of
simulations address the solution of the fluid equations (Smoothed
Particle Hydrodynamics, SPH) and the dynamic creation of point masses
to replace collapsed gas fragments in a similar manner. In the {\it
Delgado-Donate et al.} calculations, SPH simulations follow the
gas-dominated stage during $\approx0.5$~Myr before switching to an
N-body integration followed until the stability of most of the
multiples could be guaranteed. Each set of initial conditions is run
10 to 20 times, varying only the spectrum of the turbulent velocity
field imposed initially, in order to obtain statistically significant
average properties.

Finally, some numericists focused on the largest scales, and tried to
study the collapse and fragmentation of large clouds ($100$ to a few
$1000 M_\odot$; {\it MacLow and Klessen}, 2004, {\it Padoan and
Nordlund}, 2002). These models reproduce the filamentary structure
observed in molecular clouds, and find that cores are not quasi-static
structures, but rather grow in mass by accretion and merge
hierarchically until a specific core mass function (resembling the
initial mass function at the high mass end) is built. While these
simulations provide an important step for our understanding of the
collapse of entire molecular clouds, they have little, if any,
predictive power regarding the properties of individual or multiple
stars. We do not consider these simulations in this review, which
focusses on testing models against some of the most constraining
observational data available, the properties of multiple stars at the
earliest stages. These simulations are discussed in other
chapters to this volume ({\it Ballesteros-Paredes et al.}, {\it
  Klein et al.}).

\bigskip
\noindent
\textbf{ 4.2 Predictions And Comparisons With Observations}
\bigskip

These numerical simulations make a number of predictions regarding the
statistical properties of young multiple systems. Ideally, these
predictions should be tested against the observational results
summarized in {\it Mathieu et al.} (2000) and in previous sections of
this chapter, for instance. However, because of the limited
parameter space that has been explored to date, and the daunting task
of including all relevant physical processes, such predictions may
still be premature. In the following, we consider a few such
predictions, focusing primarily on general trends rather than detailed
quantitative predictions, and briefly mention some other models that
could be relevant for the formation of multiple systems.
\bigskip

\noindent {\it 4.2.1. Triples And Higher-Order Multiples}. The {\it
Delgado-Donate et al.} simulations produce a wealth of multiple
systems. The companion frequency (average number of companions per
primary) at $0.5$~Myr after the initiation of star formation is close
to $100\%$, whereas the frequency of multiple systems (ratio of all
binary and higher-order systems to all primaries) is $\approx 20 \%$;
in other words, for each binary/multiple system, there are 4 isolated
single objects. Clearly, multiple star formation is a major channel
for star formation in turbulent flows, as found also by {\it Bate et
al.}  (2003), {\it Bate} (2005) and {\it Goodwin et al.} (2004a,b),
but in these simulations, high-order multiples are more frequent than
binaries. The systems can adopt a variety of configurations, like
binaries orbiting binaries or triples. Such exotic systems have been
observed, and currently, the occurrence of high-order multiples among
main sequence field stars is of order $\sim15$--25\% (e.g., {\it
Tokovinin}, 2004). A similar proportion of multiple systems was found
in both T~Tauri and embedded protostars populations (see Section~2.5
and 3.2).  The {\it Goodwin et al.} (2004b) calculations,
characterized by a very low ratio of initial turbulent energy to
thermal energy, produce a significantly lower number of stars per
core, and match better the observed multiplicity fractions. This has
led {\it Goodwin \& Kroupa} (2005) to propose that the main mode for
star formation involves the break-up of a core into 2 to 3 stars, a
larger number being a rare outcome.  This is a revealing constraint on
star formation theories although it does not settle the question of
how this main mode of 2-to-3 fragments comes to be (low turbulence is
a possibility but there may be others) and which multiplicity
properties we should expect from it.  \bigskip

\noindent {\it 4.2.2. Multiplicity As A Function Of Age}. {\it
Delgado-Donate et al.} found that the companion frequency decreases
during the first few Myr of N-body evolution, as many of the initial
multiple systems are unstable. It must be noted that, although the
canonical timescale for dynamical break-up of an unstable multiple is
at most $10^5$ yr, this timescale is significantly increased in these
gas dynamical simulations because of two effects. First, numerous
companions form on orbits with very large separations, up to orbital
periods approaching the $10^5$ yr timescale. Second, it is found in
gas dynamical calculations that star formation always occurs in
bursts, which repeat themselves with decreasing intensity for several
cluster free-fall times, i.e. several $\times$ $10^5$ yr, adding new
stars to already formed, and maybe already stable, multiples. While
the former effect may be a somewhat artificial consequence of the
selected initial conditions, the latter is very robust, and likely
applies to most star formation scenarios.

The total companion frequency is seen to rapidly decay from $\approx
100 \%$ to $\approx 30\%$. This internal decay affects mostly low-mass
outliers, which are released in vast amounts to the field. It might be
expected that in a real cluster the companionship would drop even
further -- or sooner -- as star forming cores do not form in isolation
but close to one another. Weakly bound outliers might have been
stripped sooner by torques from other cluster members (see simulations
by {\it Kroupa et al.}, 1999, for instance). The total frequency of
binary and high-order multiples, on the other hand, varies little
after the first few $\times$ $10^5$~yr. Thus, although a multiple
system is still likely to evolve further towards its hierarchical
stable configuration in longer timescales, the relative frequency of
singles, binaries, triples, and so on, seems to be essentially
established after a few $\times 10^5$~yr.

Based on these models, one may expect that the frequency of
binary/high-order multiples among T~Tauri stars, which are already a
few Myr old, should be essentially the same as that of main sequence
field stars. While this is true for young clusters, loose associations
clearly show a higher companion frequency, in disagreement with this
prediction. It must be emphasized that the simulations discussed here
consider prestellar cores as isolated entities, whereas core-core
interactions may play an important role in shaping the outcome of the
star formation process. Nonetheless, to reconcile these simulations
with the observations, one can argue that a large number of the single
objects produced by the simulations are not included in the observed
samples of T~Tauri stars, either because they are of too low-mass or
because they have already been expelled from the molecular cloud owing
to their high ejection velocities (see below) and the shallow
potential well of the cloud. Both explanations have their own
weaknesses: on one hand, the ratio of stellar-mass objects to brown
dwarfs is on order 4 in the Taurus molecular cloud ({\it Guieu et
al.}, 2006) and low-mass stars are frequently members of binary and
multiple systems (see Section~2.2), and on the other hand, the much
younger embedded protostars -- although they may already be too old to
be compared with the simulations -- do not appear to have a
dramatically lower proportion of single stars (see Section~3.2).
\bigskip

\noindent {\it 4.2.3. Multiplicity As A Function Of Primary Mass}. The
{\it Delgado-Donate et al.} models find a positive dependence of the
multiplicity fraction on primary mass (see Fig.~4), in qualitative
agreement with observations. The low- and high- mass end of the
distribution, however, do not match satisfactorily the observations;
the dependence on mass is too steep. The models fail to produce as
many low-mass binaries as observed because of their extremely low
binding energy, making them prone to disruption in an environment
dominated by dynamical interactions. Observationally, a binary
fraction of at least $15\%$ is seen among field brown dwarfs
(e.g. {\it Bouy et al.}, 2003, {\it Mart\'{\i}n et al.}, 2003), and
values as high as $30-40\%$ have been suggested ({\it Maxted and
Jeffries}, 2005; but see {\it Joergens}, 2006). Secondly, the
high-mass end shows a paucity of singles. This is a common outcome of
all `turbulent' star formation simulations to date, and stems from the
fact that the most massive members of the cluster are always in
binaries and thus, the binary fraction at the high-mass end is close
to 100\%, whereas even in the Taurus-Auriga cloud, there are a number
$\gtrsim1M_\odot$ T~Tauri stars that have no known
companion. Calculations by {\it Goodwin et al.}  also find similar
problems to fit the observed multiplicity dependence of primary mass.
It is likely that by simulating the evolution of an ensemble of clouds
with different initial masses, thus following the successful
prescription of a 2-step procedure pioneered by {\it Durisen et al.}
(2001), the problem would be lessened, but it is unclear at the moment
-- until parameter space is more widely investigated -- whether it
would solve it completely or not. The problem of the formation of a
significant population of low-mass/brown dwarf binaries would remain,
and it has been suggested that initial conditions less prone to
fragmentation or resolution effects may provide a possible solution to
this riddle ({\it Clarke and Delgado-Donate}, in prep.). As mentioned
before, {\it Durisen et al.} (2001) and {\it Sterzik and Durisen}
(2003) extensively discuss how a steep multiplicity vs mass
correlation can be smoothed, and manage to do so by means of their
2-step mass selection.

\begin{figure}[htb]
\epsscale{0.75} \plotone{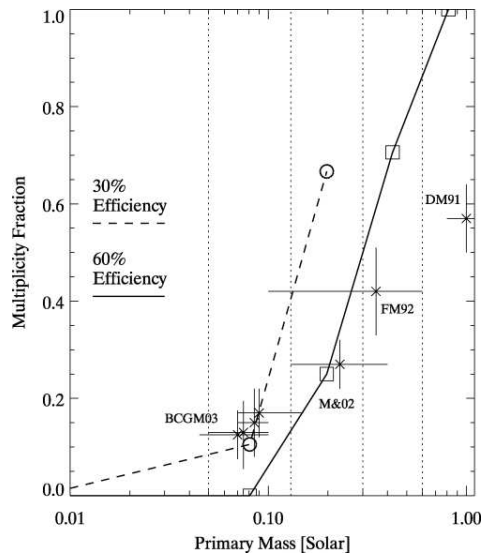}
\caption{\small Predicted mass-dependence of the multiplicity rate
  (solid and dashed lines), from {\it Delgado-Donate et al.}
  (2004b). Observational datapoints are for field stars ({\it
  Duquennoy and Mayor}, 1991; {\it Fischer and Marcy}, 1992; {\it
  Marchal et al.}, 2003; {\it Bouy et al.}, 2003, in order or
  decreasing primary mass).}
\end{figure}
\bigskip

\noindent {\it 4.2.4. Velocity Dispersion}. The simulations show that
single and binary stars attain comparable velocities in the range
$1-10$~km~s$^{-1}$, whereas higher-order multiples display lower
velocity dispersions. This kinematic segregation as a function of $N$
is the expected outcome of the break-up of unstable multiples, whereby
the ejected objects (typically singles, or less often binaries)
acquire large velocities, whereas the remaining more massive multiple
recoils with a lower speed. Among the singles and binaries, the peak
of the velocity distribution is of order a few km~s$^{-1}$, in the
range of the cloud random velocities. A similar velocity distribution
is produced by N-body models ({\it Sterzik and Durisen}, 2003) and
models with lower levels of turbulence ({\it Goodwin et al.},
2004a,b). Therefore, we would expect low-mass star-forming regions
like Taurus, where a local kinematic segregation may survive against
the influence of large scale dynamics, to display an overabundance of
multiple systems in the densest regions, from where the high speed
low-mass singles would have escaped. This prediction was made by {\it
Delgado-Donate et al.}  (2004b), and has been recently supported by
the simulations of {\it Bate} (2005).

On the observational side, we note that the radial velocity outliers
found by {\it Covey et al.} (2005) and discussed in Section~3.2 could be
ejected protostars; only a long-term monitoring of their radial
velocities will help determine their status. From another perspective,
the most recent survey for low-mass Taurus members by {\it Guieu et
  al.}  (2006), covering several times the area of previous surveys,
has found that the fraction of brown dwarfs increases as one moves
away from the densest cores, known before to be over-abundant in
binaries. This could be an indication for an average larger speed for
the lower-mass single objects, as predicted. A complete analysis of
the spatial distribution of multiple systems in the Taurus cloud has
not been performed yet, but would provide a crucial test of this
prediction. Furthermore, there are some caveats on whether the census
of low-mass stars in the extended area covered by {\it Guieu et al.}
(2006) is complete. More observational work is still required to test
this critical prediction of numerical simulations.  \bigskip

\noindent {\it 4.2.5. Other models}. So far, we have reviewed models
based on the pure N-body break-up of small clusters and models where
an `initial' turbulent velocity field is imposed to the cloud, so that
its decay triggers the formation of structure until the Jeans
instability takes over and produce multiple fragmentation. However,
there are researchers that advocate a less dynamic view of star
formation. This more quiescent star formation mode may be thought of
as an extension of the {\it Shu et al.} (1987) paradigm of single star
formation to multiple systems, whereby a core in quasi-static
equilibrium collapses from the inside out in such a way that only a
few independent fragments are formed. The statistical properties of
multiple systems formed in this way are almost impossible to predict
in the absence of a detailed physical framework for this ``quiescent
fragmentation'', but can be constrained {\it a posteriori}. To
simultaneously match the high frequency of companions to low-mass
stars and the paucity of quadruple and higher-order multiples among
populations of T~Tauri stars and embedded protostars, {\it Goodwin and
Kroupa} (2005) have concluded that this star formation mode must
produce primarily binary and triple systems.  {\it Goodwin et al.}'s
low turbulence simulations are the closest we have at the moment to a
paradigm of not-so-dynamic star formation.  Alternatively, {\it
Sterzik et al.} (2003) also find that clusters with low N are a better
match for current observations.

In addition, there exists the possibility that the numerical scheme
used in most models reviewed in this section, i.e. SPH, may not
perform entirely satisfactorily in some of the regimes modeled,
especially when shear flows or voids are involved. There are
alternatives to SPH, most of them based on adaptive mesh refinement
techniques, that could offer a different view on the problem. Efforts
by Padoan and collaborators go in that direction, as well as those by
Klein and coworkers, but the complexity of the codes and the
implementation of sink particles or their equivalent for grid codes,
essential to follow star formation calculations beyond the formation
of the first star, have proved a serious obstacle so far to produce
simulations comparable in predictive power to the SPH ones. In
addition, the role of feedback in the star formation process, e.g.
through outflows, has never been included in such simulations, although
it may be important. The effects of photoionizing feedback by massive
stars have been preliminarily studied by {\it Dale et al.} (2005), who
find that photoionization fronts may have both a positive and
negative effect in star formation, by triggering fragmentation and
collapse at the HII fronts, or disrupting incoming accretion flows
respectively. These new developments are likely to shed light on some
of the issues over which theory stays the furthest apart from
observations.  \bigskip

\noindent{\textbf{ 5. CONCLUSIONS AND PERSPECTIVES}} 
\bigskip

Binary and higher-order multiple systems represent the preferred
outcome of the star formation process; studying the statistical
properties of these systems at various evolutionary stages, therefore,
offers indirect constraints on the core fragmentation and on the
subsequent dynamical interaction between gas and stars. While the
(high) frequency of Myr-old T~Tauri stars has now been long
established, there has been tremendous progress in recent years: new
populations of young stars have been surveyed, the frequency of
high-order multiple and spectroscopic binaries among T~Tauri stars is
being accurately estimated, and embedded protostars have for the first
time been surveyed for multiplicity. In the meantime, numerical
simulations describing the fragmentation and dynamical evolution of
prestellar cores towards fully-formed stars and multiple systems have
made tremendous progress, and while they may not yet allow for a fine
comparison of predictions with observations, they already predict
significant trends that can be tested.

All these studies have provided important clues towards the star
formation process, but a number of open questions remain to be solved.
For instance, the apparent uniformity of the multiplicity rate of
embedded protostars independent of environment is quite puzzling given
the strong dependence to initial conditions of all numerical
simulations of core fragmentation. Another surprising observational
result is the existence of a fairly large proportion of low-binding
energy multiple systems, which rarely survive the violent early
star-and-gas dynamical evolution in numerical simulations of collapse
and fragmentation. The absence of aggregates of more than 4--5 stars
on scales of a couple thousand AUs is also surprising, as they seem to
be ubiquitous in numerical simulations of the fragmentation and
collapse of gas clouds. Could we be missing a number of young stars in
molecular clouds, completely biasing our multiplicity surveys? If so,
then we may wonder how useful the traditional Class 0--I--II--III
evolutionary sequence really is: if stars usually form as part of
unstable multiple systems, then many stars probably ``jump'' from one
category to another over very short timescales, which could have
dramatic consequences for their circumstellar environments. The
existence of systems pairing stars of different evolutionary
categories (including the ``infrared companion'' systems among T~Tauri
stars, which are not discussed in this chapter) could be footprints of
this violent evolution, and would deserve increased attention in
upcoming years.

To investigate these and many other issues described above, continuing
both survey and follow-up efforts related to the multiplicity of young
stars appears as a crucial endeavor for the future. While high-angular
resolution ground-based infrared methods are bound to provide
important new results, one must also remember that radio
interferometric observations have the potential to complement infrared
surveys in two major ways: first, by allowing the study of the
youngest and most embedded protostars so far inaccessible to other
techniques, and secondly, by providing images of extremely tight
systems, which are so far only known to be binaries because of
spectroscopy.  They may also offer opportunities to examine the
results of few-body disruption of initially non-hierarchical
systems. In parallel to these observational efforts, more numerical
simulations must be run to sample a wider parameter space than has
been currently explored, and a general effort to include as many
physical effects as possible must be undertaken in upcoming years to
allow for direct comparisons of simulations with observations.
\bigskip

\centerline\textbf{ REFERENCES}
\bigskip
\parskip=0pt {\small \baselineskip=11pt 
\refs Allen L. E., Myers P. C., Di Francesco J., Mathieu R., Chen
H., and Young E. (2002) {\it Astrophys. J., 566}, 993-1004.

\refs Andr\'{e} P. and Montmerle T. (1994) {\it Astrophys. J.,
420}, 837-862.

\refs Anosova J. P. (1986) {\it Astrophys. and Space Science, 124},
217-241.

\refs Artymowicz P. and Lubow S. H. (1996) {\it
  Astrophys. J., 467}, L77-L80.

\refs Baraffe I., Chabrier G., Allard F., and Hauschildt
P. H. (2002) {\it Astron. Astrophys., 382}, 563-572.

\refs Barsony M., Greene T. P., and Blake G. A. (2002) {\it
Astrophys. J., 572}, L75-L78.

\refs Bate M. R. (2005) {\it Mon. Not. Roy. Astron. Soc., 363},
363-378 .

\refs Bate M. R. and Bonnell I. A. (1997) {\it
  Mon. Not. Roy. Astron. Soc., 285}, 33-48.

\refs Bate M. R., Bonnell I. A., and Price N. M. (1995) {\it
  Mon. Not. Roy. Astron. Soc., 277}, 362-376.

\refs Bate M. R., Bonnell I. A., and Bromm V. (2002) {\it
  Mon. Not. Roy. Astron. Soc., 336}, 705-713.

\refs Bate M. R., Bonnell I. A., and Bromm V. (2003) {\it
  Mon. Not. Roy. Astron. Soc., 339}, 577-599.

\refs Beck T. L., Schaefer G. H., Simon M., Prato L., Stoesz
J. A., and Howell R. R. (2004) {\it Astrophys. J., 614}, 235-251.

\refs Berger E., Ball S., Becker K. M., Clarke M., Frail D. A.,
Fukuda T. A., Hoffman I. M., Mellon R., Momjian E., Murphy N. W.,
Teng S. H., Woodruff T., Zauderer B. A., and Zavala R. T. (2001) {\it
  Nature, 410}, 338-340.

\refs Bonnell I. A., Bate M. R., Clarke C. J., and Pringle
J. E. (1997) {\it Mon. Not. Roy. Astron. Soc., 285}, 201-208.

\refs Bonnell I. A., Bate M. R., Clarke C. J., and Pringle
J. E. (2001) {\it Mon. Not. Roy. Astron. Soc., 323}, 785-794.

\refs Bonnell I. A., Bate M. R., and Vine S. G. (2003) {\it
  Mon. Not. Roy. Astron. Soc., 343}, 413-418.

\refs Bouy H., Brandner W., Mart\'{\i}n E. L., Delfosse X.,
Allard F., and Basri G. (2003) {\it Astron. J., 126}, 1526-1554.

\refs Brandeker A., Jayawardhana R., and Najita J. (2003) {\it
  Astron. J., 126}, 2009-2014.

\refs Chandler C. J., Brogan C. L., Shirley Y. L., and Loinard
L. (2005) {\it Astrophys. J., 632}, 371-396.

\refs Chauvin G., M\'enard F., Fusco T., Lagrange A.-M., Beuzit
J.-L., Mouillet D., and Augereau J.-C. (2002) {\it
Astron. Astrophys., 394}, 949-956.

\refs Chauvin G., Thomson M., Dumas C., Beuzit J.-L., Lowrance
P., Fusco T., Lagrange A.-M., Zuckerman B., and Mouillet
D. (2003) {\it Astron. Astrophys., 404}, 157-162.

\refs Chick K. M. and Cassen P. (1997) in {\it Herbig-Haro flows
and the birth of stars} (B. Reipurth and C. Bertout, eds.), p. 207,
IAU Symp. 182, Kluwer, Dordrecht.

\refs Ciardi D. R., Telesco C. M., Williams J. P., Fisher R. S.,
Packham C., Pi\~{n}a R., and Radomski J.  (2003) {\it
Astrophys. J., 585}, 392-397.

\refs Ciardi D. R., Telesco C. M., Packham C., G\'{o}mez Martin
C., Radomski J. T., De Buizer J. M., Phillips C. J., and Harker
D. E. (2005) {\it Astrophys. J., 629}, 897-902.

\refs Covey K. R., Greene T. P., Doppmann G. W., and Lada
C. J. (2006) {\it Astron. J., 131}, 512-519.

\refs Covino E., Catalano S., Frasca A., Marilli E., Fern\'andez
M., Alcal\'a J. M., Melo C., Paladino R., Sterzik M. F., and
Stelzer B. (2000) {\it Astron. Astrophys., 361}, L49-L52.

\refs Covino E., Frasca A., Alcal\'a J. M., Paladino R., and
Sterzik M. F. (2004) {\it Astron. Astrophys., 427}, 637-649.

\refs Curiel S., Girart J. M., Rodr\'{i}guez L. F., and Cant\'{o}
J.  (2003) {\it Astrophys. J., 582}, L109-L113.

\refs Dale J. E., Bonnell I. A., Clarke C. J., and Bate
M. R. (2005) {\it Mon. Not. Roy. Astron. Soc., 358}, 291-304.

\refs Delgado-Donate E. J., Clarke C. J., and Bate M. R. (2003)
{\it Mon. Not. Roy. Astron. Soc., 342}, 926-938.

\refs Delgado-Donate E. J., Clarke C. J., and Bate M. R. (2004a)
{\it Mon. Not. Roy. Astron. Soc., 347}, 759-770.

\refs Delgado-Donate E. J., Clarke C. J., Bate M. R., and Hodgkin
S. T. (2004b), {\it Mon. Not. Roy. Astron. Soc., 351}, 617-629.

\refs Duch\^ene G., Monin J.-L., Bouvier J., and M\'enard
F. (1999) {\it Astron. Astrophys., 351}, 954-962.

\refs Duch\^ene G., Ghez A. M., McCabe C., and Weinberger
A. J. (2003) {\it Astrophys. J., 592}, 288-298.

\refs Duch\^ene G., Bouvier J., Bontemps S., Andr\'e P., and
Motte F. (2004) {\it Astron. Astrophys., 427}, 651-665.

\refs Dulk G. A. (1985) {\it Ann. Rev. Astron. Astrophys., 23},
169-224.

\refs Duquennoy A. and Mayor M. (1991) {\it Astron. Astrophys.,
248}, 485-524.

\refs Durisen R. H., Sterzik M. F., and Pickett B. K. (2001) {\it
  Astron. Astrophys., 371}, 952-962.

\refs Feigelson E. D. and Montmerle T. (1999) {\it
  Ann. Rev. Astron. Astrophys., 37}, 363-408.

\refs Fischer D. A. and Marcy G. W. (1992) {\it Astrophys. J.,
  396}, 178-194.

\refs Furlan E., Forrest W. J., Watson D. M., Uchida K. I.,
Brandl B. R., Keller L. D., and Herter T. L. (2003) {\it
  Astrophys. J., 596}, L87-L90.

\refs Gammie C. F. (2001) {\it Astrophys. J., 553}, 174-183.

\refs Ghez A. M., Neugebauer G., and Matthews K. (1993) {\it
  Astron. J., 106}, 2005-2023.

\refs Ghez A. M., Weinberger A. J., Neugebauer G., Matthews K.,
and McCarthy D. W. Jr. (1995) {\it Astron. J., 110}, 753-765.

\refs Girart J, M., Rodr\'{i}guez L. F., and Curiel S. (2000) {\it
Astrophys. J., 544}, L153-L156.

\refs Girart J.~M., Curiel S., Rodr{\'{\i}}guez L.~F., Honda M.,
Cant{\'o} J., Okamoto Y.~K., and Sako S. (2004) {\it Astron. J.,
127}, 2969-2977.

\refs G\'omez L., Rodr\'{\i}guez L. F., Loinard L., Lizano S.,
Poveda A., and Allen C. (2006) {\it Astrophys. J.}, in press.

\refs Goodwin S. P. and Kroupa P. (2005) {\it Astron. Astrophys.,
439}, 565-569.

\refs Goodwin S. P., Whitworth A. P., and Ward-Thompson D. (2004a)
      {\it Astron. Astrophys., 414}, 633-650.

\refs Goodwin S. P. Whitworth A. P., and Ward-Thompson D. (2004b)
      {\it Astron. Astrophys., 423}, 169-182.

\refs Greene T. P., Wilking B. A., Andr\'{e} P., Young E. T., and
Lada C. J. (1994) {\it Astrophys. J., 434}, 614-626.

\refs G\"udel M. (2002) {\it Ann. Rev. Astron. Astrophys., 40},
217-261.

\refs Guieu S., Dougados C., Monin J.-L., Magnier E., and
Mart\'{\i}n E. L. (2006) {\it Astron. Astrophys., 446}, 485-500.

\refs Haisch Jr., K. E., Barsony M., Greene T. P., and Ressler
M. E. (2002) {\it Astron. J., 124}, 2841-2852.

\refs Haisch Jr., K. E., Greene T. P., Barsony M., and Stahler
S. W. (2004) {\it Astron. J., 127}, 1747-1754.

\refs Haisch Jr., K. E., Barsony M., Greene T. P., and Ressler
M. (2006) {\it Astron. J.}, submitted.

\refs Halbwachs J. L., Mayor M., Udry S., and Arenou F. (2003)
      {\it Astron. Astrophys., 397}, 159-175.

\refs Hamaguchi K., Corcoran M. F., Petre R., White N. E.,
Stelzer B., Nedachi K., Kobayashi N., and Tokunaga A. T. (2005)
{\it Astrophys. J., 623}, 291-301.

\refs Hartigan P. and Kenyon S.J. (2003) {\it Astrophys. J., 583},
334-357.

\refs Hearty T., Fern\'andez M., Alcal\'a J. M., Covino E., and
Neuh\"auser R. (2000) {\it Astron. Astrophys., 357}, 681-685.

\refs Hillenbrand L. A. and White R. J. (2004) {\it Astrophys. J.,
  604}, 741-757.

\refs Jensen E. L. N., Mathieu R. D., and Fuller G. A. (1994) {\it
  Astrophys. J., 429}, L29-L32.

\refs Joergens V. (2006) {\it Astron. Astrophys., 446}, 1165-1176.

\refs Johnston K. J., Fey A. L., Gaume R. A., Hummel C. A.,
Garrington S., Muxlow T., and Thomasson P. (2004) {\it
  Astrophys. J., 604}, L65-L69.

\refs Johnston K. J., Fey A. L., Gaume R. A., Claussen M. J., and
Hummel C. A. (2005) {\it Astron. J., 128}, 822-828.

\refs Kiseleva L. G., Eggleton P. P., and Mikkola S. (1998) {\it
Mon. Not. Roy. Astron. Soc., 300}, 292-302.

\refs Konopacky Q. M., Ghez A. M., McCabe C., Duch\^ene G., and
Macintosh B. A. (2006) {\it Astron. J.}, submitted.

\refs Koresko C. D. (2000) {\it Astrophys. J., 531}, L147-L149.

\refs Koresko C. D. (2002) {\it Astron. J., 124}, 1082-1088.

\refs Kraus A. L., White R. J., and Hillenbrand L. A. (2005) {\it
  Astrophys. J., 633}, 452-459.

\refs Kroupa P. (1995) {\it Mon. Not. Roy. Astron. Soc., 277},
1491-1506.

\refs Kroupa P., Petr M. G., and McCaughrean M. J. (1999) {\it New
Astron., 4}, 495-520.

\refs Lada C. J. (1987) in {\it Star Forming Regions}, (M. Peimbert
and J. Jugaku, eds.), pp. 1-17, Reidel, Dordrecht.

\refs Leinert Ch., Zinnecker H., Weitzel N., Christou J., Ridgway
S. T., Jameson R., Hass M., and Lenzen R. (1993) {\it
  Astron. Astrophys., 278}, 129-149.

\refs Lodato G. and Rice W. K. M. (2005) {\it
  Mon. Not. Roy. Astron. Soc., 358}, 1489-1500.

\refs Loinard L. (2002) {\it Rev. Mex. Astron. Astrofis., 38}, 61-69.

\refs Loinard L., Rodr\'{\i}guez L. F., D'Alessio P., Wilner
D. J., and Ho P. T. P. (2002) {\it Astrophys. J., 581},
L109-L113.

\refs Loinard L., Rodr\'{\i}guez L. F., and Rodr\'{\i}guez
M. I. (2003) {\it Astrophys. J., 587}, L47-L50.

\refs Loinard L., Mioduszewski A. J., Rodr\'{\i}guez L. F.,
Gonz\'alez R. A., Rodr\'{\i}guez M. I., and Torres R. M. (2005)
{\it Astrophys. J., 619}, L179-182.

\refs Looney L. W., Mundy L. G., and Welch W. J. (2000) {\it
Astrophys. J., 529}, 477-498.

\refs Luhman K. L. (2001) {\it Astrophys. J., 560}, 287-306.

\refs Mac Low M.-M. and Klessen R. S. (2004) {\it Rev. Modern
  Phys., 76}, 125-194.

\refs Marchal L., Delfosse X., Forveille T., S\'egransan D.,
Beuzit J.-L., Udry S., Perrier C., and Mayor M. (2003) in {\it
  Brown dwarfs} (Mart\'{\i}n, ed.), p. 311, IAU Symp. 211,
Astron. Soc. Pac., San Francisco.

\refs Mart\'{\i}n E. L., Barrado y Navascu\'es D., Baraffe I.,
Bouy H., and Dahm S. (2003) {\it Astrophys. J., 594}, 525-532.

\refs Mathieu R. D. (1994) {\it Ann. Rev. Astron. Astrophys., 32},
465-530.

\refs Mathieu R. D., Ghez A. M., Jensen E. L. N., and Simon
M. (2000) in {\it Protostars and Planets IV} (V. Mannings, A. P. Boss
and S. S. Russell, eds.), p. 703-730, Univ. of Arizona Press, Tucson.

\refs Maxted P. F. L. and Jeffries R. D. (2005) {\it
  Mon. Not. Roy. Astron. Soc., 362}, L45-L49.

\refs McCabe C., Ghez A. M., Prato L., Duch\^ene G., Fisher
R. S., and Telesco C. (2006) {\it Astrophys. J., 636}, 932-951.

\refs McDonald J. M. and Clarke C. J. (1995) {\it
  Mon. Not. Roy. Astron. Soc., 275}, 671-684.

\refs Meeus G., Sterzik M., Bouwman J., and Natta A. (2003) {\it
  Astron. Astrophys., 409}, L25-L29.

\refs Melo C. H. F. (2003) {\it Astron. Astrophys., 410}, 269-282.

\refs Noriega-Crespo A., Moro-Martin A., Carey S., Morris P. W.,
Padgett D. L., Latter W. B., and Muzerolle J. (2004) {\it
Astrophys. J. Suppl., 154}, 402-407.

\refs Ochi Y., Sugimoto K., and Hanawa T. (2005) {\it
  Astrophys. J., 623}, 922-939.

\refs Osterloh M. and Beckwith S. V. W. (1995) {\it Astrophys. J.,
  439}, 288-302.

\refs Padoan P. and Nordlund A. (2002) {\it Astrophys. J.,
  576}, 870-879.

\refs Patience J. and Duch\^ene G. (2001) in {\it The formation of
  binary stars} (H. Zinnecker and R. D. Mathieu, eds.), pp. 181-190,
  IAU Symp. 200, Astron. Soc. Pac., San Francisco.

\refs Phillips R. B., Lonsdale C. J., Feigelson E. D., and Deeney
B. D. (1996) {\it Astron. J., 111}, 918-929.

\refs Prato L. and Simon M. (1997) {\it Astrophys. J., 474},
455-463.

\refs Prato L., Greene T.P., and Simon M. (2003) {\it
  Astrophys. J., 584}, 853-874.

\refs Reipurth B. (2000) {\it Astron. J. 120}, 3177-3191.

\refs Reipurth B. and Zinnecker H. (1993) {\it Astron. Astrophys.,
278}, 81-108.

\refs Reipurth B., Rodr\'{\i}guez L. F., Anglada G., and Bally
J. (2002) {\it Astron. J., 124}, 1045-1053.

\refs Reipurth B., Rodr\'{\i}guez L. F., Anglada G., and Bally
J. (2004) {\it Astron. J., 127}, 1736-1746.

\refs Ressler M. E. and Barsony M. (2001) {\it Astron. J., 121},
1098-1110.

\refs Rodr\'{\i}guez L. F. (1997) in {\it Herbig-Haro flows and the
  birth of stars} (B. Reipurth and C. Bertout, eds.), p. 83, IAU
  Symp. 182, Kluwer, Dordrecht.

\refs Rodr\'{\i}guez L. F., D'Alessio P., Wilner D. J., Ho
P. T. P., Torrelles J. M., Curiel S., G\'omez Y., Lizano S.,
Pedlar A., Cant\'o J., and Raga A. C. (1998) {\it Nature, 395},
355-357.

\refs Rodr\'{\i}guez L. F., Porras A., Claussen M. J., Curiel S.,
Wilner D. J., and Ho P. T. P. (2003) {\it Astrophys. J., 586},
L137-L139.

\refs Rodr\'{\i}guez L. F., Loinard L., D'Alessio P., Wilner
D. J., and Ho P. T. P. (2005a) {\it Astrophys. J., 621},
L133-L136.

\refs Rodr\'{\i}guez L. F., Poveda A., Lizano S., and Allen
C. (2005b) {\it Astrophys. J., 627}, L65-L68.

\refs Shu F. H., Adams F. C., and Lizano S. (1987) {\it
  Ann. Rev. Astron. Astrophys., 25}, 23-81.

\refs Sigalotti L. Di G. and Klapp J. (2001) {\it Astron. Astrophys.,
  378}, 165-179.

\refs Stassun K. G., Mathieu R. D., Vaz L. P. R., Stroud N., and
Vrba F. J.. (2004) {\it Astrophys. J. Suppl., 151}, 357-385.

\refs Steffen A. T., Mathieu R. D., Lattanzi M. G., Latham D. W.,
Mazeh T., Prato L., Simon M., Zinnecker H., and Loreggia
D. (2001) {\it Astron. J., 122}, 997-1006.

\refs Sterzik M. F. and Durisen R. H. (1995) {\it
  Astron. Astrophys., 304}, L9-L12.

\refs Sterzik M. F. and Durisen R. H. (1998) {\it
  Astron. Astrophys., 339}, 95-112.

\refs Sterzik M. F. and Durisen R. H. (2003) {\it
  Astron. Astrophys., 400}, 1031-1042.

\refs Sterzik M. F. and Tokovinin A. A. (2002) {\it
  Astron. Astrophys., 384}, 1030-1037.

\refs Sterzik M. F., Durisen R. H., and Zinnecker H. (2003) {\it
  Astron. Astrophys., 411}, 91-97.

\refs Sterzik M. F., Melo C. H. F., Tokovinin A. A., and van der
Bliek N. (2005) {\it Astron. Astrophys., 434}, 671-676.

\refs Symington N. H., Harries T. J., Kurosawa R., and Naylor
T. (2005) {\it Mon. Not. Roy. Astron. Soc., 358}, 977-984.

\refs Tamazian V. S. (2004) {\it Astron. J., 127}, 2378-2381.

\refs Tamazian V. S., Docobo J. A., White R. J., and Woitas
J. (2002) {\it Astrophys. J., 578}, 925-934.

\refs Tan J. C. (2004) {\it Astrophys. J., 607}, L47-L50.

\refs Tohline J. E. (2002) {\it Ann. Rev. Astron. Astrophys., 40},
349-385.

\refs Tokovinin A. A. (2004) {\it Rev. Mex. Astron. Astrofis., 21},
7-14.

\refs Tokovinin A. A. and Smekhov M. G. (2002) {\it
  Astron. Astrophys., 382}, 118-123.

\refs Tsujimoto M., Kobayashi N., and Tsuboi Y. (2005) {\it
Astron. J., 130}, 2212-2219.

\refs Valenti J. A. and Johns-Krull C. M. (2004) {\it
  Astrophys. and Space Science, 292}, 619-629.

\refs Welty A.D. (1995) {\it Astron. J., 110}, 776-781.

\refs Whitney B. A., Wood K., Bjorkman J. E., and Wolff M. J. (2003)
  {\it Astrophys. J., 591}, 1049-1063.

\refs Woitas J., K\"ohler R., and Leinert Ch. (2001) {\it
  Astron. Astrophys., 369}, 249-262.

\refs Wooten A. (1989) {\it Astrophys. J., 337}, 858-864.

\refs Zuckerman B. and Song I. (2004) {\it
Ann. Rev. Astron. Astrophys., 42}, 685-721.

}

\end{document}